\documentclass[preprint,aps,draft,showpacs]{revtex4}

\begin{document}

\title{Dissipative Dynamics of a Josephson Junction  In the Bose-Gases}
\author{R.A. Barankov,$^{1}$ S.N. Burmistrov$^{2}$}
\affiliation{$^1$Department of Physics,  Massachusetts Institute of
Technology, Cambridge, Massachusetts 02139\\
$^2$RRC ``Kurchatov Institute", Kurchatov Sq.1, 123182, Moscow, Russia}

\date{\today}
\begin{abstract}
The dissipative dynamics of a Josephson junction in the Bose-gases is
considered within the framework of the model of a tunneling Hamiltonian. The
effective action which describes the dynamics of the phase difference across
the junction is derived using functional integration method. The dynamic
equation obtained for the phase difference across the junction is analyzed for
the finite temperatures in the low frequency limit involving the radiation
terms. The asymmetric case of the Bose-gases with the different order
parameters is calculated as well.
\end{abstract}
\maketitle
\section{Introduction}
The experimental realization of Bose-Einstein condensation
in atomic vapors~\cite{Anderson95,Davis95,Mewes96,Bradley97} has allowed to
observe a great variety of macroscopic quantum effects. In particular, there
arises a considerable interest to the study of the Josephson effect in the
Bose-condensed gases as one of intriguing possibilities to explore the
macroscopic quantum effects related directly to the broken symmetry in the
quantum systems. The dynamics of the Josephson effect is governed by the
difference between the phases  of the condensates, playing a role of
macroscopic quantum variable.

The theoretical treatment of the Josephson effect includes both the internal
effect for atoms of a gas in the different  hyperfine states and the case of
the Bose-condensates spatially separated with a potential barrier which acts as
a tunneling junction. The latter case due to its direct analogy with
superconductors seems us more attractive. A lot of work  has already been done
in this direction. In~\cite{Dalfovo96} the behavior of the  condensate density
near the potential boundary has been discussed and  the quasiclassical
expression for the current through a potential barrier has been obtained. The
articles~\cite{Jack96,Steel98} are devoted to an  applicability of the two-mode
approximation in the Josephson junction dynamics.  Milburn~{\it et al}
in~\cite{Milburn97} have shown an existence of the self-trapping effect as well
as the collapse and revival sequence in the  relative population.
In~\cite{Smerzi97,Raghavan99,Smerzi00} the nonlinear Josephson dynamics and
macroscopic fluctuations have been considered, resulting in the optimum
conditions~\cite{Williams01} to observe the Josephson  oscillations.
Zapata~{\it et al}~\cite{Zapata98} have presented a semiclassical  description
of the Josephson junction dynamics. The time-dependent variational  analysis of
the Josephson effect is given in~\cite{Lin00}.

One of the most interesting and important aspects in the Josephson junction
dynamics from both the theoretical and the experimental viewpoints is the
dephasing of the Josephson oscillations due to coupling between the macroscopic
relative phase variable and the infinite number of the microscopic degrees of
freedom ~\cite{Villain99,Meier01}. Historically, in the case of the
superconducting systems such description of the phase dynamics was developed in
the middle of 1980's~\cite{Amb82,Larkin83,Eckern84}. The most important result
was a successive  derivation of the effective action for the relative phase,
revealing the key role of the microscopic degrees of freedom in the
irreversible dynamics of the superconducting Josephson junctions. From the
mathematical point of view the response functions in the effective action,  which prove to be nonlocal in time,  give the full information on the dynamics
of a junction. The employment of the low frequency expansion for the response
functions allows one to obtain the dissipative dynamics of a superconducting
junction, involving Josephson energy,  renormalization of the junction capacity
(inverse effective mass), and resistance (effective friction) of a junction.

For the system of two Bose-condensates connected with a weakly coupled junction,
it is very desirable to trace and explore the dynamics of the relative phase,
generalizing the method of the derivation of the effective action from the
superconducting case to the case of the Bose-condensed systems. As we will show
in the next sections, the gapless sound-like spectrum of low energy excitations
in the Bose-condensed gases results in a qualitative change of the irreversible
phase dynamics compared with that of the superconducting junctions. So, the
main aim of the paper is to derive the effective action for the Bose point-like
junction within the framework of the functional integration method in order to
find the explicit expressions for the response functions and analyze the low
frequency dynamics of a Bose junction.

The plan of the article is the following. First, we derive the general
expression for the effective action depending only on the relative phase for
the system of two Bose-condensates connected by a point-like junction. Then we
consider the case of zero temperature. As a next step, we investigate the
effect of finite temperature on the phase dynamics. In addition, from the low
frequency expansion of the response functions we find the Josephson energy,
renormalization of the effective mass, friction coefficient, and the radiation
corrections. The latter can be interpreted as a sound emission from the
region of a Bose junction. Finally, we present the case of an asymmetric
junction in the Appendix and summarize the results in the Conclusion.

\section{Effective action}
First, it may be useful to make some remarks on the geometry of the Bose
junction and condensates. We keep in mind the case of a point or weakly coupled
junction due to a large potential barrier between the two macroscopic infinite
reservoirs containing Bose-condensates. So, we can neglect the feed-back effect
of the junction on the Bose-condensates and assume that the both condensates
are always in the thermal equilibrium state with the constant density depending
on the temperature alone. The traditional image of such system is two bulks
with one common point through which the transmission of particles is only
possible with some tunneling amplitude.

So, our starting point is the so-called tunneling Hamiltonian ($\hbar=1$,
volume  $V=1$)
\begin{equation}\label{ham}
H=H_{l}+H_{r}+H_{u}+H_{t},
\end{equation}
where $H_{l,r}$ describes the bulk Bose-gas on the left-hand and right-hand
sides, respectively,
\begin{equation}
H_{l,r}=\int d^{3}r~\Psi_{l,r}^{+}\left( -\frac{\Delta}{2m}-\mu
+\frac{u_{l,r}}{2}\Psi_{l,r}^{+}\Psi_{l,r}\right) \Psi_{l,r}.
\end{equation}
The  coupling constant $u_{l,r}=4 \pi a_{l,r}/m$ where as usual $a_{l,r}$ is
the scattering length.
The energy
\begin{equation}
H_{u}=\frac{U}{2}\left( \frac{N_{l}-N_{r}}{2}\right) ^{2},
\end{equation}
is analogous to the capacity energy of a junction in the case of
superconductors. The constant $U$ can be associated with the second derivative
of the total energy $E=E(N_l, N_r)$ with respect to the relative change in
the number of particles across the junction
\begin{equation} U= \left(\frac{\partial
^{2} }{\partial N_l^2 }+\frac{\partial ^{2}}{\partial N_r^2 }\right)E,
\end{equation}
and usually is estimated as $U=(\partial\mu _{l}/\partial N_{l} +\partial\mu
_{r}/\partial N_{r})$~\cite{Zapata98}. In general, it may depend on the
concrete type of the Bose-junction and simply describes that the energy of the
system on the whole may depend on the relative number of particles from each
bulk. The total number of the particles in each bulk is given by
\begin{equation}
N_{l,r}=\int d^{3}r~\Psi_{l,r}^{+}\Psi_{l,r}.
\end{equation}
The term
\begin{equation}
H_{t}=-\int\limits_{r\in l,r^{\prime}\in
r}d^{3}rd^{3}r^{\prime}~\left[\Psi_{l}^{+}\left( {\bf r}\right) I\left( {\bf
r,r^{\prime}}\right)\Psi_{r}\left({\bf r^{\prime}}\right) +h.c. \right]
\end{equation}
is responsible for the transitions of particles from the right-hand to the
left-hand bulk and {\it vice versa}.

To study the properties of the system described by~(\ref{ham}), we calculate
the partition function using the analogy of the superconducting junction
\begin{equation}
Z=\int {\cal D}^2 \Psi _{l}{\cal D}^2 \Psi _{r}~\exp \left[ -S_E\right],
\end{equation}
where the action on the Matsubara (imaginary) time reads
\begin{equation}
\begin{array}{c}
S_E=\int\limits_{-\beta /2}^{\beta /2}d\tau ~L_E,\\
L_E=\int d^{3}r\left\{ \Psi _{l}^{+}~\frac{\partial }{\partial \tau }~\Psi
_{l}+\Psi _{r}^{+}~\frac{\partial }{\partial \tau }~\Psi _{r}\right\} + H.
\end{array}
\end{equation}
To eliminate the quartic term in the action which comes from the
``capacity" energy $H_{u}$, we use the Hubbard-Stratonovich procedure by
introducing an additional gauge field $V(\tau)$ on the analogy with the
so-called plasmon gauge field in metals
\begin{equation}
\begin{array}{c}
\exp \left[ -\frac{U}{2}\int d\tau ~\left( \frac{N_{l}-N_{r}}{2}\right) ^{2}
\right] =
\int {\cal D} V\exp \left\{ -\int d\tau ~\left[ \frac{V^{2}\left(
\tau \right) }{2U}+i\frac{N_{l}-N_{r}}{2}V\left( \tau \right) \right]
\right\},\\
\int {\cal D}V\exp \left[ -\int d\tau ~\frac{V^{2}\left( \tau \right)
 }{2U}\right]=1.
\end{array}
\end{equation}

Next, we follow the Bogoliubov method of separating the field operators into the
condensate and non-condensate fractions, i.e., $\Psi_{l,r}=c_{l,r}+\Phi_{l,r}$.
Denoting $x=(\tau,{\bf r})$ and introducing convenient Nambu spinor notations
for the field operators and, correspondingly,  matrices for the Green functions
and tunneling amplitudes, we arrive at the following expression for the
partition function
\begin{equation}
Z=\int {\cal D}V{\cal D}^2C\exp \left[ -S_{0}\right] \int {\cal
D}^2\Phi \exp\left[ -S_{\Phi }\right].\\
\end{equation}

Thus, we split the initial action into the two parts which correspond to the
condensate and noncondensate fields, respectively,
\begin{equation}\label{action}
\begin{array}{l}
S_{0}=\int d\tau \left(
\begin{array}{l}
c_{l}^{+}\left( \frac{\partial }{\partial \tau }-\mu
+i\frac{V\left( \tau \right) }{2}\right) c_{l}+\frac{u_{l,r}}{2}
c_{l}^{+}c_{l}^{+}c_{l}c_{l}+(l\rightarrow r, V\rightarrow -V)\\
-I_{0}\left(
c_{l}^{+}c_{r}+c_{r}^{+}c_{l}\right) +\frac{V^{2}\left( \tau \right) }{2U}
\end{array}
\right),\\
\\
S_{\Phi }=\int dxdx^{\prime
}\left\{ \Phi ^{+}\left( \widetilde{G}^{\left( 0\right)
-1}-\widetilde{I}\right) \Phi -C^{+}\widetilde{I}\Phi -\Phi
^{+}\widetilde{I}C\right\}.
\end{array}
\end{equation}

For the field operators we used the spinor notations
\begin{equation}\label{notation}
\Phi =\frac{1}{\sqrt{2}}\left(
\begin{array}{c}
\Xi _{l} \\
\Xi _{r}
\end{array}
\right) , \,\, \Xi _{l,r}=\left(
\begin{array}{c}
\Phi _{l,r} \\
\Phi _{l,r}^{+}
\end{array}
\right), \,\,
C=\frac{1}{\sqrt{2}}\left(
\begin{array}{c}
C_{l} \\
C_{r}
\end{array}
\right), \,\,
C_{l,r}=\left(
\begin{array}{c}
c_{l,r} \\
c_{l,r}^{+}
\end{array}
\right).
\end{equation}

In the expression for the condensate part $S_{0}$ we define the amplitude
$I_{0}$ equal to
\begin{equation}
I_0 = \int d^3 r\, d^3 r^{\prime}\, I ({\bf r}, {\bf r^{\prime}}).
\end{equation}
and corresponding to the tunneling process of the condensate-to-condensate
particles.  It is straightforward to obtain the following expressions for the
matrix Green functions
\begin{equation}
\widetilde{G}^{\left( 0\right) -1}=\left(
\begin{array}{cc}
\widehat{G}_{l}^{\left( 0\right) -1} & 0 \\
0 & \widehat{G}_{r}^{\left( 0\right) -1}
\end{array}
\right) \delta \left( x-x^{\prime }\right),
\end{equation}

\begin{equation}
\widehat{G}_{l,r}^{\left( 0\right) -1}=\left(
\begin{array}{cc}
G_{l,r}^{\left( 0\right) -1}+\Sigma _{11}^{l,r} & \Sigma _{20}^{l,r} \\
\Sigma _{02}^{l,r} & \overline{G}_{l,r}^{\left( 0\right) -1}+\overline{\Sigma
} _{11}^{l,r}
\end{array}
\right),
\end{equation}
where the inverse Green functions and self-energy parts are given by the
well-known expressions
\begin{equation}\label{Green}
\begin{array}{l}
G_{l,r}^{\left( 0\right) -1}=\frac{\partial }{\partial \tau }-\frac{\Delta }{
2m}-\mu \pm i\frac{V\left( \tau \right) }{2},\\
\\
\Sigma _{11}^{l,r}=2u_{l,r}\, c_{l,r}^{+}c_{l,r}, \,\, \Sigma
_{20}^{l,r}=u_{l,r}\, c_{l,r}c_{l,r}, \,\, \Sigma _{02}^{l,r}=u_{l,r}
\,c_{l,r}^{+}c_{l,r}^{+}.
\end{array}
\end{equation}
Accordingly, the matrix Green function $\widehat{G}^{(0)}_{l,r}$ can be
represented as
\begin{equation}
\label{GF}
\widehat{G}^{(0)}_{l,r}
=\left(
\begin{array}{cc}
G_{l,r} & F_{l,r} \\
F^{+}_{l,r} & \overline{G}_{l,r}
\end{array}
\right).
\end{equation}
The transfer matrix here has the form
\begin{equation}
\widetilde{I}
=\left(
\begin{array}{cc}
0 & \widehat{I} \\
\widehat{I}^{\ast } & 0
\end{array}
\right),\,\,\,
\widehat{I}=\left(
\begin{array}{cc}
I & 0 \\
0 & \overline{I}
\end{array}
\right),\,\,\, I=I\left(x,x^{\prime }\right) =I\left( {\bf r},{\bf r^{\prime
 }}\right)
\delta \left( \tau -\tau ^{\prime }\right).
\end{equation}

As one can readily see, if we employ a gauge transformation of the field
 operators
\begin{equation}
\Psi_{l,r} \rightarrow \exp \left[i\varphi
_{l,r}\left( \tau \right)\right] \Psi_{l,r},
\end{equation}
and impose the conditions $\dot{\varphi} _{l}=-V/2$, $\dot{\varphi} _{r}=V/2$,
i.e.,
\begin{equation}
\dot{\varphi}=V, \,\,\,
\varphi =\varphi _{r} -\varphi_{l},
\label{fidot}
\end{equation}
both normal $G_{l,r}$ and anomalous Green functions $F_{l,r}$~(\ref{GF})
gain additional phase factors with respect to the functions in the lack of an
external field (notations of~\cite{AGD}).

\begin{equation}
\begin{array}{l}
G_{l,r}\left( \tau ,\tau ^{\prime }\right) \rightarrow \exp \left( i
[\varphi _{l,r}\left( \tau \right) -\varphi _{l,r}\left( \tau ^{\prime
}\right)] \right) G_{l,r}\left( \tau -\tau ^{\prime }\right),\\
\\
F_{l,r}\left( \tau ,\tau ^{\prime }\right) \rightarrow \exp \left( i
[\varphi _{l,r}\left( \tau \right) +\varphi _{l,r}\left( \tau^{\prime }
\right) ]\right) F_{l,r}\left( \tau -\tau ^{\prime }\right).
\end{array}
\end{equation}

The part of action $S_\Phi$ in~(\ref{action}) is quadratic in the
non-condensate field operators so we can integrate them out. To perform the
integration, we employ the well-known formula
\begin{equation} \int {\cal D}^2
\Phi \exp \left[ -\Phi ^{+}\alpha \Phi +\beta ^{+}\Phi +\Phi ^{+}\beta \right]
=\exp \left[ \beta ^{+}\alpha ^{-1}\beta -\mathop{\rm Tr} \left[ \ln \left(
\alpha \right) \right] \right], \end{equation} with $\alpha = {G}^{\left(
0\right) -1}-\widetilde{I}$ and $\beta =\widetilde{I} C$ to arrive at  the
partition function \begin{equation}
Z=\int {\cal D}\varphi{\cal D}^2C\exp \left[ -S\right]
\end{equation}
with the effective action given by
\begin{equation}\label{preciseaction}
S=S_0-{\rm Tr} \left[C^{+}\widetilde{I}\left(\widetilde{G}^{\left(
0\right) -1}-\widetilde{I} \right)^{-1}\widetilde{I} C\right] +\mathop{\rm Tr}
\left[ \ln \left( \widetilde{G}^{\left( 0\right) -1}-\widetilde{I} \right)
\right]. \end{equation}

In order to run analytically further, it is necessary to make the
following  approximations. First, we expand the second and third terms
of~(\ref{preciseaction}) in powers of the tunneling amplitude $I$  to the
first nonvanishing order. Then, as is stated above, we consider the simplest
case of a point-like junction putting $I\left( x,x^{\prime }\right)
=I_{0}\delta \left(  {\bf r}\right) \delta \left( {\bf r^{\prime }}\right)
\delta \left( \tau -\tau ^{\prime }\right)$. The latter also allows us to escape
from the problem of  summing all higher-order terms in the tunneling amplitude
$I$, which is inherent in a  junction of the plane
geometry~\cite{Meier01,Babich01} with the conservation of  the tangential
components of the momentum of a tunneling particle. The problem  in essence
becomes one-dimensional \cite{Babich01,Gesh01} and  results in a strongly
dissipative low-frequency  dynamics independent of the tunneling amplitude and
governed by the bulk  relaxation alone. In our consideration this would
correspond to the amplitude  independent on the $x$ and $y$ coordinates, i.e.,
$I\propto \delta (z)\delta (z^{\prime})$. Finally, the third approximation we
use is a saddle-point approximation for the condensate part of the partition
function.  Substituting $c_{l,r}=\sqrt{n_{0 l,r}}$ where $n_{0 l,r}$ is the
density of particles in the condensate fraction, we obtain the expression for
the partition function depending on the phase difference alone
\begin{equation}
Z_{\varphi}=\int {\cal D} \varphi ~\exp \left(
-S_{eff}\left[ \varphi \right]
\right).
\end{equation}
The corresponding effective action reads
\begin{equation}\label{effaction}
\begin{array}{l}
 S_{eff}\left[ \varphi \right] =\int d\tau
 ~\left[ \frac{1}{2U}\left( \frac{d\,\varphi }{d\,\tau }\right)
 ^{2}-2I_{0}\sqrt{n_{0 l}n_{0 r}}\cos \varphi \right] - \\
I_{0}^{2}\int d\tau d\tau ^{\prime }~\left\{
  \begin{array}{l}
    \alpha \left( \tau -\tau ^{\prime
       }\right) \cos \left[ \varphi \left( \tau \right) -\varphi \left( \tau
       ^{\prime }\right) \right] \\
           +\beta \left( \tau -\tau ^{\prime }\right)
        \cos  \left[ \varphi \left( \tau \right) +\varphi \left( \tau ^{\prime
         }\right)  \right]
   \end{array}
 \right\}.
\end{array}
\end{equation}
Here the response functions can be written using the Green functions
\begin{equation}\label{response}
\begin{array}{l}
\alpha \left( \tau \right) =n_{0r}g^{+}_l\left( \tau \right)+
n_{0 l}g^{+}_r\left( \tau \right) +G\left(\tau\right),\\
\\
\beta \left( \tau \right) =n_{0 r}f_l\left( \tau \right) +n_{0 l}f_r\left(
\tau \right) +F\left(\tau\right),
\end{array}
\end{equation}
where
\begin{equation}
\begin{array}{l}
g^{\pm}_{l,r}\left( \tau \right) =\int \frac{d^{3}p}{2\left( 2\pi \right) ^{3}}
\left[G_{l,r}\left( p,\tau \right)\pm G_{l,r}\left( p,-\tau \right)\right],\,
f_{l,r}\left( \tau \right) =\int \frac{d^{3}p}{\left( 2\pi \right) ^{3}}
F_{l,r}\left( p,\tau \right),\\
\\
G\left(\tau\right)=2\left[g^{+}_l\left( \tau \right)g^{+}_r\left( \tau \right)
-g^{-}_l\left( \tau \right) g^{-}_r\left( \tau \right)\right],\,
F\left(\tau\right)=2f_l\left(\tau\right) f_r\left(\tau \right).
\end{array}
\end{equation}
The Fourier components of the Green functions of a weakly
interacting Bose-gas are given by~\cite{AGD}
\begin{equation}\label{bosegreen}
\begin{array}{l}
G_{l,r}\left( p,\omega _{n}\right) =\frac{i\omega _{n}+\xi
_{p}+\Delta_{l,r}}{\omega _{n}^{2}+\varepsilon_{l,r} ^{2}\left( p\right) },\,
F_{l,r}\left( p,\omega _{n}\right) =-\frac{\Delta_{l,r} }{\omega
_{n}^{2}+\varepsilon_{l,r} ^{2}\left( p\right) },\\
\\
\varepsilon_{l,r} ^{2}\left( p\right) =\xi _{p}^{2}+2\Delta_{l,r} \xi
_{p},\quad \xi _{p}=\frac{ p^{2}}{2m}\,\, , \quad \Delta _{l,r}=u_{l,r}n_{0
l,r}. \end{array}
\end{equation}
Thus, in order to comprehend the dynamics of the relative phase difference
$\varphi$ across the junction, one should analyze the behavior of the response
functions $\alpha$ and $\beta$ as a function of time.

\section{The response functions.}
The calculation of the response functions in the general form is a rather
complicated problem. However, keeping in view, first of all, the study of the
low frequency dynamics  of a junction,  we can restrict our calculations by
analyzing the behavior of  the response functions on the long-time scale. This
means that we should find the low frequency decomposition of the response
functions in the Matsubara frequencies. Next, we will use the  procedure of
analytical continuation in order to derive the dynamic equation  which the
relative phase $\varphi$ obeys.

\subsection{Fourier transformation of the $\alpha$-response.}
From the analysis of the Green function behavior $g^{\pm}_{l,r}(\tau )$ it
follows that  zero Fourier component of the $\alpha$-response function
diverges. The simplest way to avoid this obstacle is to deal with the difference
$\tilde\alpha\left( \omega_{n} \right) =\alpha \left( \omega_{n}
\right)-\alpha (0)$. This corresponds to the substitution
$\alpha(\tau)=\tilde\alpha(\tau)+\alpha(0)\delta(\tau)$ into effective
action~(\ref{effaction}) and the second term $\alpha (0)\delta (\tau )$ yields
a physically unimportant time-independent contribution into the action, meaning
a shift of the ground state energy of a junction. Following this procedure, we
can arrive at the explicit expressions of the Fourier components for all the
terms in~(\ref{response}).

For the first two terms in the $\alpha$-response function of~(\ref{response}),
we have a simple formula
\begin{equation}
\tilde g^{+}\left( \omega _{n}\right) =-\frac{\pi \nu }{\sqrt{2} }
\left[ \sqrt{1+\left|\frac{\omega _{n}}{\Delta} \right|}-1\right],
\end{equation}
and the corresponding expansion  to  third order in $\omega /\Delta $ is
\begin{equation}
\tilde g^{+}\left( \omega _{n}\to 0\right)
\approx -\frac{\pi \nu }{2\sqrt{2}}\left|\frac{\omega _{n}
}{\Delta }\right|\left[ 1-\frac{1}{4}\left|\frac{\omega _{n}}{\Delta
}\right|+\frac{1}{8}\left| \frac{\omega _{n}}{\Delta }\right|^{2}-\ldots
\right].
\end{equation}
Here, $\nu =m\sqrt{ m\Delta}/(\sqrt{2}\pi ^{2})$ is the density of states
at the energy equal to $\Delta$ in the normal gas.

In the case of $\tilde G$ the
calculations of the Fourier components are much more complicated. So, we could
find the expressions only in the case of zero temperature and the first
nonvanishing order in temperature. Note that nonzero temperature effects are
connected with the behavior of $\tilde G(\omega)=\tilde G_0(\omega)+\tilde
G_T(\omega)$.

For zero-temperature part of $\tilde G(\omega)$, we obtain
\begin{equation}\label{gzero}
\begin{array}{l}
\tilde G_0\left( \omega\right) =-\nu _{l}\nu _{r}\Delta _{l}\Delta _{r}\omega
^{2}\int\limits_{0}^{\infty }\int\limits_{0}^{\infty }\frac{dxdy\;\sqrt{
(\sqrt{x^{2}+1}-1)(\sqrt{y^{2}+1}-1)}}{\left[ \left(
\Delta _{l}x+\Delta _{r}y\right) ^{2}+\omega ^{2}\right] \left( \Delta
_{l}x+\Delta _{r}y\right) }\\
\left(
1-\frac{xy}{\sqrt{\left(x^{2}+1\right)\left(y^{2}+1\right)}}\right).
\end{array}
\end{equation}

Unfortunately, we could not evaluate expression~(\ref{gzero}) in the explicit
analytic form. Thus we report here the Fourier expansion up to third order
in $\omega$
\begin{equation}
\tilde G_0\left( \omega\to 0\right) =-\frac{\pi\nu _{l}\nu _{r}\omega
^{2}}{8\sqrt{\Delta_l\Delta_r}}\left[\phi\left(\frac{\Delta_l-\Delta_r}{\Delta_l
+ \Delta_r}\right)-
\frac{1}{3}\frac{|\omega|}{\sqrt{\Delta_l\Delta_r}}+\dots\right].
\end{equation}
Here we introduced the function
\begin{equation}
\label{dilog}
\begin{array}{r}
\phi(q)=\frac{2\pi}{3}+\frac{1}{q^2}\left(1-3
q^2-\sqrt{1-q^2}\right)-\frac{2}{\pi}\ln^2\left[\sqrt{\frac{1+q}{1-q}}\right]-\\
\frac{2}{\pi}\left(
\begin{array}{l}
{\rm Li}_2\left[-\sqrt{\frac {1+q}{1-q}}\right] +{\rm Li}_2\left[-\sqrt{\frac
{1-q}{1+q}}\right]+\\
{\rm Li}_2\left[1-\sqrt{\frac {1+q}{1-q}}\right]+{\rm
Li}_2\left[1-\sqrt{\frac {1-q}{1+q}}\right]
\end{array}
\right),
\end{array}
\end{equation}
where ${\rm Li_2(z)}=\int^{0}_z d\,t \ln(1-t)/t$ is the dilogarithm function.
The Taylor expansion of $\phi(q)$ in the case of $|q|\ll 1$ reads:
\begin{equation}
\phi(q)\approx\pi-\frac{5}{2}+\frac{q^2}{8}+\ldots
\end{equation}

In the case of the equal order parameters $\Delta=\Delta_l=\Delta_r$ we obtain:
\begin{equation}
\tilde G_0\left(\omega \to
0\right) \approx -\frac{\pi \nu ^{2} \Delta}{8
}\left|\frac{\omega}{\Delta}\right|^{2} \left[ \pi
-\frac{5}{2}-\frac{1}{3}\left|\frac{\omega }{\Delta }\right|+\ldots \right].
\end{equation}
The other finite-temperature contribution into $\tilde G$ can readily be
evaluated as
\begin{equation}
\tilde G_T\left( \omega \right) =\frac{\pi \nu _{l}\nu _{r}\sqrt{\Delta
_{l}\Delta _{r}}}{24}\left( \frac{2\pi T}{\sqrt{\Delta _{l}\Delta _{r}}}\right)
^{2}\left[ \sqrt{\frac{\Delta _{l}}{\Delta _{r}}}+\sqrt{\frac{\Delta _{r}}{
\Delta _{l}}}-\sqrt{\frac{\Delta _{l}+|\omega _{n}|}{\Delta _{r}}}-\sqrt{
\frac{\Delta _{r}+|\omega _{n}|}{\Delta _{l}}}\right]+\dots,
\end{equation}
which in the case of $\Delta=\Delta_l=\Delta_r$ gives:
\begin{equation}\label{gnonzero}
\tilde G_T\left( \omega \right) =\frac{\pi \nu ^2\Delta}{12}\left(
\frac{2\pi T}{\Delta }\right) ^{2}\left[
1-\sqrt{1+\left|\frac{\omega_n}{\Delta}\right|}\right]+\dots.
\end{equation}
Expanding this expression to third order in $\omega _n /\Delta$ and including
zero temperature terms yields
\begin{equation}\label{lzero}
\tilde G\left(\omega _{n}\to 0\right) \approx -
\frac{\pi \nu^2 \Delta }{24}\left|\frac{\omega _{n}}{\Delta}\right|\left[
\begin{array}{l}
\left(\frac{2\pi T}{\Delta }\right) ^{2}+\left|\frac{\omega
_{n}}{\Delta}\right|\left[3\pi -\frac{15}{2}-\frac{1}{4}\left(\frac{2\pi
T}{\Delta }\right) ^{2}\right]-\\
\left|\frac{\omega_{n}}{\Delta}\right|^2 \left[1-\frac{1}{8}\left(\frac{2\pi
T}{\Delta }\right) ^{2}\right]-\ldots
\end{array}
\right].
\end{equation}

It is worthwhile to emphasize the appearance a linear term in $|\omega_{n}|$ in
Eq.(\ref{lzero}) at finite temperatures. As we will see below, this results in
 the
manifestation of an additional temperature-dependent contribution into the
dissipation of the junction. The effect can be interpreted as a tunneling of
normal thermal excitations existing in the system due to finite temperatures.

\subsection{Fourier transformation of the $\beta$-response.}
Like the preceding section, we can evaluate the Fourier components and
find the low frequency expansion for the anomalous Green functions $f(\tau)$.
For the first two terms in the $\beta$-response function~(\ref{response}), we
obtain
\begin{equation}
f\left( \omega _{n}\right) =-\frac{\pi \nu }{\sqrt{2}\sqrt{
1+|\omega _{n}/\Delta |}}
\mathrel{\mathop{\approx }\limits_{\omega _{n} \to 0}}
-\frac{\pi \nu }{\sqrt{2} }\left( 1-\frac{1}{2}\left|\frac{\omega _{n}}{
\Delta }\right|+\frac{3}{8}\left|\frac{\omega _{n}}{\Delta
}\right|^{2}-\frac{5}{16}\left|\frac{ \omega _{n}}{\Delta }\right|^{3}+\ldots
\right).
\end{equation}
The calculation of $F(\omega_n)$ requires a special attention. In fact,
analyzing the formula of the Fourier transform
\begin{equation}
F\left( \omega _{n}\right) =\frac{\pi ^{2}\nu _{l}\nu _{r}\sqrt{\Delta
_{l}\Delta _{r}}}{\beta}\sum\limits_{k=-\infty}^{\infty }\frac{1}{
\sqrt{\left(\Delta _{l}+|\omega _{k}|\right)\left(\Delta
_{r}+|\omega _{k}-\omega _{n}|\right)}},
\end{equation}
we see that the zero-temperature expression
\begin{equation}
F\left( \omega \right) =\frac{\pi \nu _{l}\nu _{r}\sqrt{\Delta _{l}\Delta _{r}}}{2
}\int\limits_{-\infty }^{\infty }d\omega
^{\prime }\frac{1}{\sqrt{\left(\Delta _{l}+|\omega ^{\prime
}|\right)\left(\Delta _{r}+|\omega ^{\prime }-\omega |\right)}},
\end{equation}
diverges logarithmically at $\omega=0$ due to behavior of the integrand at
large frequencies  $\omega^{\prime}\to\infty$. To perform integration over
$\omega^{\prime}$, we should first pay attention that so far we could neglect
the dependence on the momentum in the self-energy parts since all the integrals
gain their values in the region of small momenta. In fact, the self-energy
parts depend on the momentum and we must use the exact expressions $\Sigma
_{20}\left(p\right) =\Sigma_{02}\left( p\right) =n_{0}U\left( p\right) $ where
$U\left( p\right) $ is a Fourier component of the interaction between
particles, $\Sigma _{20}\left(0\right)$ being $n_{0}U\left( 0\right) = 4\pi a
n_{0}/m$. Here $a$ is the scattering length, $n_{0}$ is the particle density,
and $m$ is the mass of a particle. The order of the magnitude for the momentum
at which $U\left( p\right)$ decays can be estimated roughly as a reciprocal of
the scattering length, i.e., $\ p\simeq 1/a$. So, we put the upper limit of
integration equal to the cutoff frequency $\omega_{c}\simeq
1/ma^{2}$ within the logarithmic accuracy.

Finally, representing $F(\omega _n)$ as a sum of zero-temperature and finite
temperature contributions $F(\omega_n)=F_0(\omega_n)+F_T(\omega_n)$,
we find the zero-temperature term
\begin{equation}
F_0\left( \omega _{n}\right)
=\frac{\pi \nu _{l}\nu _{r}\sqrt{\Delta _{l}\Delta _{r}}}{2}\left(
\begin{array}{c}
2\ln \left[ \frac{4\omega _{c}}{\left( \sqrt{\Delta _{l}}+\sqrt{\Delta _{r}}
\right) ^{2}}\right] -2\ln \left[ \frac{\left( \sqrt{\Delta _{l}}+\sqrt{\Delta
_{r}+|\omega _{n}|}\right) \left( \sqrt{\Delta _{r}}+\sqrt{\Delta
_{l}+|\omega _{n}|}\right) }{\left( \sqrt{\Delta _{l}}+\sqrt{\Delta _{r}}
\right) ^{2}}\right] + \\
+\arctan \frac{|\omega _{n}|+\Delta _{l}-\Delta _{r}}{2\sqrt{\left( |\omega
_{n}|+\Delta _{l}\right) \Delta _{r}}}+\arctan \frac{|\omega _{n}|+\Delta
_{r}-\Delta _{l}}{2\sqrt{\left( |\omega _{n}|+\Delta _{r}\right) \Delta _{l}}}
\end{array}
\right).
\end{equation}
and  the finite temperature one, respectively
\begin{equation}
F_T\left( \omega _{n}\right) =\frac{\pi \nu _{l}\nu _{r}\sqrt{\Delta
_{l}\Delta _{r}}}{24}\left( \frac{2\pi T}{\sqrt{\Delta
_{l}\Delta _{r}}}\right) ^{2}\left(
\begin{array}{l}
\frac{\Delta _{l}}{\sqrt{\Delta _{r}\left( \Delta _{l}+|\omega
_{n}|\right) }}+\frac{\Delta _{r}}{\sqrt{\Delta _{l}\left( \Delta
_{r}+|\omega _{n}|\right) }}\\
-\frac{\Delta _{l}+\Delta _{r}}{2\sqrt{\Delta
_{l}\Delta _{r}}}
\end{array}
\right)+\dots.
\end{equation}
These functions in the case of $\Delta_l=\Delta_r=\Delta$ take the form:
\begin{equation}
F_0\left( \omega _{n}\right)
=\pi \nu^2\Delta\left(
2\ln \left[ \frac{2\sqrt{\omega
_{c}/\Delta}}{1+\sqrt{1+|\omega_n/\Delta|}}\right] +\arctan
\frac{|\omega _{n}/\Delta|}{2\sqrt{1+|\omega _{n}/\Delta|}} \right),
\end{equation}
\begin{equation}
F_T\left( \omega _{n}\right) =\frac{\pi\nu^2\Delta}{12}\left(
\frac{2\pi T}{\Delta}\right) ^{2}\left( \frac{1}{\sqrt{1+|\omega
_{n}/\Delta| }}-\frac{1}{2}\right)+\dots.
\end{equation}

Expanding these expressions to third order in $\omega _n /\Delta$ yields
\begin{equation}\label{fzero}
F\left(\omega _{n}\to 0\right) \approx \frac{\pi
\nu^2\Delta}{24} \left(
\begin{array}{l}
24\ln \left[ \frac{\omega _{c}}{\Delta}
\right] +\left( \frac{2\pi T}{\Delta}\right) ^{2}-\left|\frac{\omega
_{n}}{\Delta }\right|\left(
\frac{2\pi T}{\Delta }\right) ^{2}- \\
\left|\frac{\omega _n}{\Delta
}\right|^2\left[ \frac{3}{2}-\frac{3 }{4}\left(
\frac{2\pi T}{\Delta } \right) ^{2}\right] + \\
\left|\frac{\omega _{n}}{\Delta}\right|^3
\left[ 1-\frac{5 }{8}\left( \frac{2\pi T}{\Delta}\right)
^{2}\right] -\dots
\end{array}
\right).
\end{equation}
Here, as in the case of $\alpha$-response we emphasize the
appearance a linear term in $|\omega_{n}|$ in Eq. (\ref{fzero}) at finite
temperatures which contributes to the dissipation and interpreted as a tunneling
of normal thermal excitations existing in the system at finite temperatures.

\subsection{Functional series of the response functions.}
The Fourier components of the response functions in the form of a series in
$|\omega_n|$ up to third order can be written in the form
\begin{equation}\label{abresp}
\begin{array}{l}
\tilde\alpha \left( \omega_{n} \right)
=-\alpha_{1}|\omega_{n}|+\alpha_{2}|\omega_{n}|^{2}-\alpha_{3}
|\omega_{n}|^{3}+\dots,\\
\\
\beta \left( \omega_{n} \right)
=-\beta_{0}+\beta_{1}|\omega_{n}|-\beta_{2}|\omega_{n}|^{2}+
\beta_{3}|\omega_{n}|^{3}+\dots.
\end{array}
\end{equation}
Accordingly,  the expressions for the response functions in the imaginary
time representation read as
\begin{equation}
\begin{array}{l}
\tilde\alpha (\tau) =\alpha_{1}\frac{1}{\pi}\left(\frac{\pi T}{\sin(\pi T
 \tau)}\right)^{2}-\alpha_{2}\delta^{\prime\prime}(\tau)-\alpha_{3}\frac{2
 (\pi T)^{4}}{\pi}\left[\frac{3}{\sin^{4}(\pi T \tau)}-\frac{2}{\sin^{2}(\pi T
 \tau)}\right]+\dots,\\
 \\
\beta (\tau) =-\beta_{0}\delta(\tau)-\beta_{1}\frac{1}{\pi}\left(\frac{\pi
 T}{\sin(\pi T \tau)}\right)^{2}+\beta_{2}\delta^{\prime\prime}(\tau)+
\beta_{3}\frac{2 (\pi T)^{4}}{\pi}\left[\frac{3}{\sin^{4}(\pi T
 \tau)}-\frac{2}{\sin^{2}(\pi T \tau)}\right]+\dots.
 \end{array}
\end{equation}
For the sake of brevity, we present expressions for $\alpha_i(T)$ and
$\beta_i(T)$ in the symmetric case of $\Delta_l=\Delta_r=\Delta$. The general
case of $\Delta_l\ne\Delta_r$ will be considered in the Appendix.
\begin{equation}
\begin{array}{l}
\alpha_{1}=\frac{\gamma}{2}\left\{1+\frac{1}{3}\sqrt{\frac{na^{3}}{\pi}}
\left (\frac{2\pi T}{\Delta}\right)^{2}\right\},\\
\alpha_{2}=\frac{\gamma}{8
 \Delta}\left\{1-4\sqrt{\frac{na^{3}}{\pi}}\left[\pi-\frac{5}{2}-\frac{1}{12}
 \left(\frac{2\pi T}{\Delta}\right)^{2}\right]\right\},\\
\alpha_{3}=\frac{\gamma}{16
 \Delta^{2}}\left\{1-\frac{8}{3}\sqrt{\frac{na^{3}}{\pi}}\left[1-\frac{1}{8}
 \left(\frac{2\pi T}{\Delta}\right)^{2}\right]\right\},\\
\beta_{0}=\gamma\Delta\left\{1-4\sqrt{\frac{na^{3}}{\pi}}\left[\ln(\frac{1}{
na^{3}})+\frac{1}{24}\left(\frac{2\pi T}{\Delta}\right)^{2}\right]\right\},\\
\beta_{1}=\frac{\gamma}{2}\left\{1-\frac{1}{3}\sqrt{\frac{na^{3}}{\pi}}\left
(\frac{2\pi T}{\Delta}\right)^{2}\right\},\\
\beta_{2}=\frac{3\gamma}{8
 \Delta}\left\{1+\frac{2}{3}\sqrt{\frac{na^{3}}{\pi}}\left[1-\frac{1}{2}
 \left(\frac{2\pi T}{\Delta}\right)^{2}\right]\right\},\\
\beta_{3}=\frac{5\gamma}{16
 \Delta^{2}}\left\{1+\frac{8}{15}\sqrt{\frac{na^{3}}{\pi}}\left[1-\frac{5}{8}
 \left(\frac{2\pi T}{\Delta}\right)^{2}\right]\right\}.
\end{array}
\end{equation}
Here $T/\Delta\ll 1$, $\gamma=\sqrt{2}\pi\nu n_0/\Delta$, $n$ is the total
density of a gas, $n_0$ is the condensate fraction density, and $a$ is the
scattering length. Note that the gas parameter $n a^3\ll 1$ naturally enters the equations.

First of all, from the Fourier expansion of $g(\omega_n)$, $f(\omega_n)$,
$G(\omega_n)$, and  $F(\omega_n)$  we conclude that the dissipation in a
point-like junction due to the presence of the linear $|\omega_n|$ term can be
associated with the various  physical processes. Thus in the case of the $g$
and $f$ contribution the dissipation can be ascribed to  the
noncondensate-condensate particle tunneling process and exists down to zero
temperature \cite{Meier01}. On the other hand, the inspection of the
expressions for $G$ and $F$ shows that these terms, producing no contribution
into dissipation at zero temperature, will be responsible for the explicit
$T^2$-behavior of the dissipative effects in the junction. The origin of this
finite temperature contribution can be attributed to the tunneling of thermal
phonon-like excitations across the junction. The meaning of the other dynamical
renormalizations and its temperature behavior will be discussed below.

\section{Josephson equation.}
To consider the dynamic behavior of the relative phase $\varphi$ in the
real time, we now follow the standard procedure of analytical continuation.
Accordingly, the substitution $|\omega_n|\rightarrow -i\omega$ in the Fourier
transform of the Euler-Lagrange equation $\delta S_{eff}/\delta\varphi (\tau)
=0$ in imaginary time entails the classical equation of motion for the relative
phase with the next inverse Fourier transformation to the real-time
representation. In particular, this means that we should replace $|\omega_n|$
with $-i\omega$ in the above expressions for the Fourier transform of the
response functions $\alpha$ and $\beta$. The effective action $S_{eff}[\varphi
(t)]$ in the real time, which variation $\delta S_{eff}/\delta\varphi (t) =0$
yields the real-time equation of motion, can be given by the expression
\begin{equation}
\begin{array}{l}
 S_{eff}\left[ \varphi \right] =\int d\,t
 ~\left[ \frac{1}{2U}\left( \frac{d\,\varphi }{d\,t }\right)
 ^{2}+2I_{0}\sqrt{n_{0 l}n_{0 r}}\cos \varphi \right] - \\
I_{0}^{2}\int d\,t d\,t ^{\prime }~\left\{
  \begin{array}{l}
    \tilde\alpha \left( t -t ^{\prime
       }\right) \cos \left[ \varphi \left( t \right) -\varphi \left( t
       ^{\prime }\right) \right] \\
           +\beta \left( t -t ^{\prime }\right)
        \cos  \left[ \varphi \left( t \right) +\varphi \left( t ^{\prime
         }\right)  \right]
   \end{array}
 \right\}.
\end{array}
\end{equation}

In the limit of the slowly varying phase the response functions in the real-time
representation can be represented in the form of a functional series.
According to Eqs.(\ref{abresp}), we have
\begin{equation}\label{dec}
\begin{array}{l}
\tilde\alpha (t)
 =\alpha_{1}\delta^{\prime}(t)-\alpha_{2}\delta^{\prime\prime}(t)+\alpha_{
3}\delta^{\prime\prime\prime}(t)+\dots,\\
\\
\beta (t)
 =\beta_{0}\delta(t)-\beta_{1}\delta^{\prime}(t)+\beta_{2}
 \delta^{\prime\prime}(t)-\beta_{3}\delta^{\prime\prime\prime}(t) +\cdots.
\end{array}
\end{equation}
Employing variational principle to the effective action
$S_{eff}[\varphi (t)]$ and using decomposition (\ref{dec}),
we can derive the Josephson equations valid for the slow variations of the
phase provided the typical time of its evolution is longer than $1/\Delta$:
\begin{equation}\label{equation}
\stackrel{\cdots}\varphi G_{3}(\varphi)+(3/2)\ddot{\varphi}\dot{\varphi}
G_{3}^{\prime}(\varphi) - \dot{\varphi}^{3}G_{3}(\varphi) +
\ddot{\varphi}G_{2}(\varphi) +(1/2)\dot{\varphi}^{2}G_{2}^{\prime}(\varphi )+
\dot{\varphi}G(\varphi) +U^{\prime}(\varphi) =0,
\end{equation}
and in accordance with Eq.(\ref{fidot})
$$
\dot{\varphi} =-\delta\mu (t)=\mu_1-\mu_2.
$$
Here we have retained the time derivatives of $\varphi (t)$ to third order
corresponding to radiation corrections.

In the following, for the sake of brevity we will consider the case of
$\Delta_l=\Delta_r=\Delta$. The general expressions for the coefficients for
$\Delta_l\ne\Delta_r$ will be considered in the Appendix.

The potential energy of a junction is given by the well-known relation
$$
U(\varphi)=-E_{J}\cos\varphi + (1/2)E_{2J}\cos 2\varphi,
$$
with the coefficients
\begin{eqnarray}
E_{J}=2 n_0 I_0,\nonumber\\
E_{2J}=G_0\Delta\left\{1-4\sqrt{\frac{na^3}{\pi}}\left[\ln\frac{1}{na^3}
+\frac{1}{24}\left(\frac{2 \pi T}{\Delta}\right) ^2\right]\right\},
\end{eqnarray} where
\begin{eqnarray}\label{conduct}
G_0=2\gamma I_0^2=16\pi \sqrt{\frac{na^3}{\pi}}(n_0 I_0/\Delta)^2,\nonumber\\
\Delta=n_0 u.
\end{eqnarray}
In the above relations the condensate density $n_0=n_0(T)$ depends explicitly on temperature according to the well-known expression for a depletion of the condensate fraction of the weakly interacting Bose gas~\cite{Fetter}. An increase of the temperature leads obviously to decreasing the Josephson energy.

The friction coefficient $G(\varphi)$ determining the Ohmic dissipation is
given by
\begin{equation}\label{friction}
G(\varphi)=G_{0}\left[\cos^{2}\varphi+\frac{1}{3} \sqrt{\frac{n
a^{3}}{\pi}}\left(\frac{2\pi T}{\Delta}\right)^{2}\sin^{2}\varphi\right].
\end{equation}
It is natural that nonzero temperature enhances the energy dissipation of a
junction due to appearance of thermal excitations in the Bose-condensed gas.

We may compare $G_0$ of the Bose-gas to the analogous conductance of
a normal Fermi gas with the same density of states:
\begin{equation}
G_N=4\pi I_0^2\nu^2(\Delta),\,\, G_N=8\sqrt{\frac{n a^3}{\pi}}G_0.
\end{equation}

The inverse effective mass of a junction is determined by
\begin{eqnarray}
G_{2}(\varphi)= U^{-1} -\alpha _2 - \beta _2 \cos 2\varphi,  \\
\alpha _2 =\frac{G_0}{8\Delta}\left\{1-4\sqrt{\frac{na^3}{\pi}}\left[
\pi -\frac{5}{2} -\frac{1}{12}\left(\frac{2\pi T}{\Delta }\right)
 ^2\right]\right\},
\nonumber \\
\beta _2 = \frac{3G_0}{8\Delta}\left\{ 1+\frac{2}{3}\sqrt{\frac{na^3}{\pi}}
\left[ 1-\frac{1}{2}\left(\frac{2 \pi T}{\Delta}\right) ^2\right]\right\}.
\nonumber \\
\end{eqnarray}
The renormalization of the effective mass results from the both
condensate-noncondensate and noncondensate-noncondensate particle tunneling
processes.

The coefficient $G_3(\varphi)$ responsible for the radiation effects reads
\begin{eqnarray}
G_3(\varphi)=\alpha _3 +\beta _3\cos 2\varphi, \nonumber\\
\alpha_{3}=\frac{G_{0}}{16\Delta^{2}}
\left\{1-\frac{8}{3}\sqrt{\frac{na^{3}}{\pi}}\left[1-\frac{1}{8}
\left(\frac{2\pi T}{\Delta}\right)^{2}\right]\right\},\\
 \beta_{3}=\frac{5G_{0}}{16
\Delta^{2}}\left\{1+\frac{8}{15}\sqrt{\frac{na^{3}}{\pi}}\left[1-
\frac{5}{8}\left(\frac{2\pi T}{\Delta}\right)^{2}\right]\right\}. \nonumber
\end{eqnarray}
These effects can be associated with emitting a sound from the
region of a junction during the tunneling of particles across the junction. On
the other hand, the radiation effects can be treated as a frequency dispersion
of the effective friction coefficient.

We will not enter here in details of the conditions which should be imposed on the
coefficients of the Josephson equation (53) in order to observe the well-defined
Josephson effect since this topic is already discussed much in the
literature. In essence, the necessary condition reduces to the requirement of
smallness either quantum zero-point or thermal fluctuations for the phase difference
across a junction, i.e., mean square  value $\langle (\Delta\varphi )^2\rangle \ll 1$.
Involving that $\langle (\Delta\varphi )^2\rangle \sim T/E_{J}$ in the thermal activation
region with the crossover at $T_0\sim E_{J}/max \{ G, \sqrt{G_{2}E_{J}} \}$
to $\langle (\Delta\varphi )^2 \rangle \sim\hbar /max\{ G, \sqrt{G{_2}E_{J}} \}$
in the quantum fluctuation regime at lower temperatures, it is desirable to have
sufficiently large Josephson energy $E_J$ or, correspondingly, not too large
potential barriers.

Another interesting aspect of such kind of experiment is an
investigation of the effects beyond the mean-field approximation
of a very dilute gas $na^3 \ll 1$, in particular, observation of
the temperature effects in the dynamics of a junction. As we have
seen, the scale of the temperature effects should reach the order
of $\sqrt{na^3}$ at $2\pi T\sim\Delta \ll T_c$. This is of much
interest since the temperature behavior of the Josephson dynamics
is closely related to the properties of elementary excitations in
a condensed Bose-gas.

The gas parameter $na^3$ under the conditions typically realized
is at most of order of 10$^{-4}$, e.g., $n\sim 10^{15}$cm$^{-3}$
and $a=5$nm for $^{87}$Rb. In principle, it is possible to
approach $na^3 \sim 1$, increasing $a\rightarrow \infty$ by
appropriate tuning of magnetic field with the Feshbach resonance.
However, too large  values of scattering length can facilitate
rapid three-body recombination. On the other hand, according to
recent work~\cite{Cornish00} such recombination may not be
inevitable at approaching $na^3 \sim 10^{-2}$. For such values of
$na^3$, the effects beyond the mean-field approximation becomes
well noticeable of about 10\%.

The second interesting aspect is connected with the analog of the
voltage-current characteristic which is an important attribute of a
superconducting junction. This kind of experiment implies maintenance
of the constant bias $\delta\mu$ for the chemical potentials across a
junction, corresponding to constant pressure drop $\delta P = mn\delta\mu$
and time dependence of the relative phase $\varphi (t) = - \delta\mu t +\varphi _0$.
One of possibilities is to use the field of Earth's gravity for this purpose.

As a result, we arrive at the following mean value of the particle current
across a Bose junction for a sufficiently large period of time as a function
of bias $\delta\mu$
\begin{equation}\label{current}
\langle I \rangle = \frac{\delta\mu}{\hbar} \langle G(\varphi )\rangle
\left[ 1- \left( \frac{\delta\mu}{\hbar}\right)^2
\frac{\langle G_3 (\varphi )\rangle}{\langle G(\varphi )\rangle }\right],
\end{equation}
where
$$
\langle G(\varphi )\rangle = \frac{1}{2} G_0 \left[ 1+\frac{1}{3}
\sqrt{\frac{na^3}{\pi}}\left(\frac{2\pi T}{\Delta }\right) ^2\right],
$$
and $\langle G_3 (\varphi )\rangle =\alpha _3$. This experiment may give
an information on the dissipative properties of a junction and also on the
nonlinear effects to be of the order of $(\delta\mu /\Delta )^3$. Note that
nonlinear effects do not contain smallness of $(na^3 )^{1/2}$.

\section{Conclusion}

To summarize, in this paper we have used a functional integration approach
for the model of a tunneling Hamiltonian in order to analyze the dynamics
of a point-like Josephson junction between two weakly non-ideal Bose gases.
The effective action and response functions which describe completely the
dynamics of a junction are found. Using the low frequency decomposition of the
response functions, the quasiclassical Josephson equation which the time
evolution of the phase difference $\varphi$ across the junction obeys is
obtained to the terms of third order in time derivatives. The corresponding
kinetic coefficients are calculated analytically, involving the
finite-temperature corrections. The temperature effect on the kinetic
coefficients demonstrates the $T^2$-behavior.

Like a junction of the planar geometry~\cite{Meier01,Babich01,Gesh01},
the dynamics of a point-like junction concerned here has an Ohmic dissipative
type due to the gapless character of excitations in a Bose-condensed gas. Note
only that the scale of the dissipative effect in the latter case is
significantly less.

The behavior of the dissipative particle current on the
temperature and the difference of the chemical potentials in
Eq.~(\ref{current}) can be used to investigate the role of
quasiparticle excitations in the dynamics of a Bose junction.
Since the dissipation is closely related to the tunneling process
of condensate particles and depends on the structure of a
junction, the effect of dissipation in the junction dynamics can
be reduced provided the structure of a junction prevents
condensate particles from tunneling across the junction. We
believe this question deserves further study.

\section{Acknowledgement}
We wish to thank V.S. Babichenko and Yu. Kagan for discussions. One of us (S.B.)
is grateful to "Statistical Physics Program" and Russian Foundation of Basic
Researches for support.

\section{Appendix}
Here we present the general case of $\Delta_l\ne\Delta_r$. To
derive expressions for the coefficients in the equation for the
phase~(\ref{equation}), we first find the Taylor expansion of the response functions
$\alpha$ and $\beta$. Finally, for the coefficients in the functional
series~(\ref{abresp}) we arrive at the expressions:
\begin{equation}
\begin{array}{l}
\alpha_{1}=\frac{\gamma}{2}\left\{b^{(2)}_{lr}+\frac{1}{3}\left(\frac{n_l
a_l^{3}n_r a_r^ { 3 } }{\pi^2}\right)^{1/4} \left (\frac{2\pi
T}{\sqrt{\Delta_l\Delta_r}}\right)^{2}\right\},\\
\alpha_{2}=\frac{\gamma}{8
\sqrt{\Delta_l\Delta_r}}\left\{b^{(3)}_{lr}-4\left(\frac{n_l
a_l^{3}n_r a_r^ { 3 } }{\pi^2}\right)^{1/4}\left[\phi_{lr} -
\frac{\Delta_l+\Delta_r} {24\sqrt{\Delta_l\Delta_r}} \left(\frac{2\pi
T}{\sqrt{\Delta_l\Delta_r}}\right)^{2}\right]\right\},\\
\alpha_{3}=\frac{\gamma}{16
\Delta_l\Delta_r}\left\{b^{(4)}_{lr}-\frac{8}{3}\left(\frac{n_l
a_l^{3}n_r a_r^ { 3 } }{\pi^2}\right)^{1/4}\left[1 -
\frac{\Delta_l^2+\Delta_r^2} {16\Delta_l\Delta_r} \left(\frac{2\pi
T}{\sqrt{\Delta_l\Delta_r}}\right)^{2}\right]\right\},\\
 \beta_0=\sqrt{\Delta_l\Delta_r}\gamma\left\{b^{(1)}_{lr}-4\left(\frac{n_l
a_l^{3}n_r a_r^ { 3 } }{\pi^2}\right)^{1/4}\left[\ln \left[ \frac{4\omega
_{c}}{\left( \sqrt{\Delta _{l}}+\sqrt{\Delta _{r}} \right) ^{2}}\right]
+\frac{\Delta _{l}+\Delta _{r}}{48\sqrt{\Delta _{l}\Delta _{r}}}\left(
\frac{2\pi T}{\sqrt{\Delta _{l}\Delta _{r}}}\right) ^{2}\right]\right\},\\
\beta_{1}=\frac{\gamma}{2}\left\{b^{(2)}_{lr}-\frac{1}{3}\left(\frac{n_l
a_l^{3}n_r a_r^ { 3 } }{\pi^2}\right)^{1/4} \left (\frac{2\pi
T}{\sqrt{\Delta_l\Delta_r}}\right)^{2}\right\},\\
\beta_{2}=\frac{3\gamma}{8
\sqrt{\Delta_l\Delta_r}}\left\{b^{(3)}_{lr}+\frac{2}{3}\left(\frac{n_l
a_l^{3}n_r a_r^ { 3 }
}{\pi^2}\right)^{1/4}\left[\frac{4\sqrt{\Delta_l\Delta_r}}{\left(\sqrt{\Delta_l}
+\sqrt{\Delta_r} \right )^2 } - \frac{\Delta_l+\Delta_r}
{4\sqrt{\Delta_l\Delta_r}} \left(\frac{2\pi
T}{\sqrt{\Delta_l\Delta_r}}\right)^{2}\right]\right\},\\
\beta_{3}=\frac{5\gamma}{16
\Delta_l\Delta_r}\left\{b^{(4)}_{lr}+\frac{8}{15}\left(\frac{n_l
a_l^{3}n_r a_r^ { 3 } }{\pi^2}\right)^{1/4}\left[1 -
\frac{5\left(\Delta_l^2+\Delta_r^2\right)} {16\Delta_l\Delta_r} \left(\frac{2\pi
T}{\sqrt{\Delta_l\Delta_r}}\right)^{2}\right]\right\}.\\
\end{array}
\end{equation}
Here we use the following notations
\begin{eqnarray}
\phi_{lr}=\phi\left(\frac{\Delta_l-\Delta_r}{\Delta_l+\Delta_r}\right),\,
\gamma=\pi\sqrt{\frac{2\nu_l\nu_r n_{0l} n_{0r}}{\Delta_l \Delta_r}},\\
b^{(n)}_{lr}=\frac{1}{2}\left[\left(\frac{\Delta_r}{\Delta_l}\right)^{n/2}\left(
\frac{ n _ l a_l^3}{n_r
a_r^3}\right)^{1/4}+\left(\frac{\Delta_l}{\Delta_r}\right)^{n/2}\left(\frac{n_r
a_r^3}{n_l a_l^3}\right)^{1/4}\right].\nonumber
\end{eqnarray}
and $\phi$ is introduced in~(\ref{dilog}).

Then, for the potential energy of a junction we have
\begin{equation}
\begin{array}{l}
E_J=2 I_0\sqrt{n_{0l} n_{0r}},\\
E_{2J}=G_0\sqrt{\Delta_l\Delta_r}\left\{b^{(1)}_{lr}-4\left(\frac{n_l
a_l^{3}n_r a_r^ { 3 } }{\pi^2}\right)^{1/4}\left[\ln \left[ \frac{4\omega
_{c}}{\left( \sqrt{\Delta _{l}}+\sqrt{\Delta _{r}} \right) ^{2}}\right]
+\frac{\Delta _{l}+\Delta _{r}}{48\sqrt{\Delta _{l}\Delta _{r}}}\left(
\frac{2\pi T}{\sqrt{\Delta _{l}\Delta _{r}}}\right) ^{2}\right]\right\}.
\end{array}
\end{equation}

$G_0$ is given by
\begin{eqnarray}
G_0=2\gamma I_0^2=16\pi\left(\frac{n_l a_l^{3}n_r a_r^ { 3 }
}{\pi^2}\right)^{1/4}\left(I_0\sqrt{\frac{n_{0l} n_{0r}}{\Delta_l \Delta_r}}\right)^2,\\
\Delta_{l,r}=n_{0 l,r} u_{l,r}.\nonumber
\end{eqnarray}
In the above relations the condensate densities $n_0=n_0(T)$ depend on temperature according to the well-known expression for the depletion of
the condensate fraction of the weakly interacting Bose gas~\cite{Fetter}.

The friction coefficient $G(\varphi)$ determining the Ohmic dissipation is
represented by
$$
G(\varphi)=G_{0}\left[b^{(2)}_{lr}\cos^{2}\varphi+\frac{1}{3}\left(\frac{n_l
a_l^{3}n_r a_r^ { 3 } }{\pi^2}\right)^{1/4} \left (\frac{2\pi
T}{\sqrt{\Delta_l\Delta_r}}\right)^{2}\sin^{2}\varphi\right].
$$
It is natural that nonzero temperature enhances the energy dissipation of a
junction due to appearance of thermal excitations in the Bose-condensed gas.

Comparing $G_0$ for the Bose-gases with the conductance of  normal
Fermi gases with the same densities of states, we have
\begin{equation}
G_N=4\pi I_0^2\nu_l\nu_r,\,G_N=8\left(\frac{n_l a_l^3 n_r
a_r^3}{\pi^2}\right)^{1/4}G_0. \end{equation}

The inverse effective mass of a junction is determined by
\begin{eqnarray}
G_{2}(\varphi)= U^{-1} -\alpha _2 - \beta _2 \cos 2\varphi,  \\
\alpha _2
=\frac{G_0}{8\sqrt{\Delta_l\Delta_r}}\left\{b^{(3)}_{lr}-4\left(\frac{n_l
a_l^{3}n_r a_r^ { 3 } }{\pi^2}\right)^{1/4}\left[\phi_{lr} -
\frac{\Delta_l+\Delta_r} {24\sqrt{\Delta_l\Delta_r}} \left(\frac{2\pi
T}{\sqrt{\Delta_l\Delta_r}}\right)^{2}\right]\right\},
\nonumber \\
\beta _2 = \frac{3G_0}{8\sqrt{\Delta_l\Delta_r}}
\left\{b^{(3)}_{lr}+\frac{2}{3}\left(\frac{n_l a_l^{3}n_r a_r^ { 3 }
}{\pi^2}\right)^{1/4}\left[\frac{4\sqrt{\Delta_l\Delta_r}}{\left(\sqrt{\Delta_l}
+\sqrt{\Delta_r} \right )^2 } - \frac{\Delta_l+\Delta_r}
{4\sqrt{\Delta_l\Delta_r}} \left(\frac{2\pi
T}{\sqrt{\Delta_l\Delta_r}}\right)^{2}\right]\right\}.
\nonumber \\ \end{eqnarray}
The renormalization of the effective mass results from the both
condensate-noncondensate and noncondensate-noncondensate particle tunneling
processes.

The coefficient $G_3(\varphi)$ responsible for the radiation effects reads
\begin{eqnarray}
G_3(\varphi)=\alpha _3 +\beta _3\cos 2\varphi, \nonumber\\
\alpha_{3}=\frac{G_{0}}{16\Delta_l\Delta_r}
\left\{b^{(4)}_{lr}-\frac{8}{3}\left(\frac{n_l a_l^{3}n_r a_r^ { 3 }
}{\pi^2}\right)^{1/4}\left[1 - \frac{\Delta_l^2+\Delta_r^2} {16\Delta_l\Delta_r}
\left(\frac{2\pi T}{\sqrt{\Delta_l\Delta_r}}\right)^{2}\right]\right\},
\\
 \beta_{3}=\frac{5
G_{0}}{16\Delta_l\Delta_r}\left\{b^{(4)}_{lr}+\frac{8}{15}\left(\frac{n_l
a_l^{3}n_r a_r^ { 3 } }{\pi^2}\right)^{1/4}\left[1 -
\frac{5\left(\Delta_l^2+\Delta_r^2\right)} {16\Delta_l\Delta_r} \left(\frac{2\pi
T}{\sqrt{\Delta_l\Delta_r}}\right)^{2}\right]\right\}.
\nonumber
\end{eqnarray}
The expression derived for the dynamical coefficients in
equation~(\ref{equation}) governing the phase difference across a point-like
junction allows us to describe the low frequency dynamics in the asymmetric
case of Bose gases with the different order parameters.

\bibliography{joseph.tex}

\begin{thebibliography}{23}
\bibitem{Anderson95}M.H. Anderson, J.R. Ensher, M.R. Matthews, C.E. Wieman,
 E.A.  Cornell, Science {\bf 269}, 198 (1995)
\bibitem{Davis95}K.B. Davis, M.-O. Mewes, M.R. Andrews, N.J. van Druten, D.S.
 Durfee,D.M. Kurn, W. Ketterle, Phys.Rev.Lett. {\bf 75}, 3969 (1995)
\bibitem{Mewes96}M.O. Mewes, M.R. Andrews, N.J. van Druten, D.M. Kurn, D.S.
 Durfee, W. Ketterle, Phys.Rev.Lett. {\bf 77}, 416 (1996)
\bibitem{Bradley97}C.C. Bradley, C.A. Sackett, R.G. Hulet, Phys.Rev.Lett. {\bf
 78},985 (1997)
\bibitem{Dalfovo96}F. Dalfovo, L. Pitaevskii, S. Stringari, Phys.Rev.A {\bf
 54}, 4213 (1996)
\bibitem{Jack96}M.W. Jack, M.J. Collett, and D.F. Walls, Phys.Rev.A {\bf 54},
R4625  (1996)
\bibitem{Steel98}M.J. Steel, M.J. Collett, Phys.Rev.A {\bf 57}, 2920 (1998)
\bibitem{Milburn97}G.J. Milburn, J. Corney, E.M. Wright, and D.F. Walls,
 Phys.Rev.A {\bf 55}, 4318(1997)
\bibitem{Smerzi97}A. Smerzi, S. Fantoni, S. Giovanazzi, S.R. Shenoy,
 Phys.Rev.Lett. {\bf 79},4950 (1997)
\bibitem{Raghavan99}S. Raghavan, A. Smerzi, S. Fantoni, S.R. Shenoy, Phys.Rev.A
 {\bf 59}, 620(1999)
\bibitem{Smerzi00}A. Smerzi, S. Raghavan, Phys.Rev.A {\bf 61}, 063601 (2000)
\bibitem{Williams01}J.E. Williams, Phys.Rev.A {\bf 64}, 013610 (2001)
\bibitem{Zapata98}I. Zapata, F. Sols, A.J. Leggett, Phys.Rev.A {\bf 57},
 R28(1998)
\bibitem{Lin00}Chi-Yong Lin, E.J.V. de Passos, Da-Shin Lee, Phys.Rev.A {\bf
 62},  055603(2000)
\bibitem{Gesh01} H.P. B\"uchler, V.B. Geshkenbein, G. Blatter, Phys.Rev.Lett.
 {\bf 87}, 100403 (2001)
\bibitem{Villain99}P. Villain, M. Lewenstein, Phys.Rev.A {\bf 59}, 2250 (1999)
\bibitem{Meier01}F. Meier, W. Zwerger, Phys.Rev.A {\bf 64}, 033610 (2001)
\bibitem{Amb82}V. Ambegaokar, U. Eckern, G. Sch\"on, Phys.Rev.Lett. {\bf 48},
 1745 (1982)
\bibitem{Larkin83}A.I. Larkin, Yu.N. Ovchinnikov, Phys.Rev.B {\bf 28}, 6281
 (1983)
\bibitem{Eckern84}U. Eckern, G. Sch\"on, V. Ambegaokar, Phys.Rev.B {\bf 30},
 6419 (1984)
 \bibitem{AGD}A.A. Abrikosov, L.P. Gorkov, I.E. Dzyaloshinski, Methods of
 Quantum Field Theory in Statistical Mechanics, Dover Publications Inc., New
 York, 1975
\bibitem{Babich01}V.S. Babichenko,  cond-mat/0109248.
\bibitem{Fetter}A. L. Fetter, J. D. Walecka, ``Quantum theory of many-particle systems", San
Francisco, McGraw-Hill, 1971
\bibitem{Cornish00}S.L. Cornish, N.R. Claussen, J.L. Roberts, E.A. Cornell, C.E. Wieman, Phys.Rev.Lett. {\bf 85}, 1795 (2000)
\end{thebibliography}

\end{document}